# Hexagonal Boron Nitride (hBN) as a Low-loss Dielectric for Superconducting Quantum Circuits and Qubits

**Authors:** Joel I-J. Wang[1†*], Megan A. Yamoah[2,3†], Qing Li[3], Amir H. Karamlou[1,3], Thao Dinh[2], Bharath Kannan[1,3], Jochen Braumueller[1], David Kim[4], Alexander J. Melville[4], Sarah E. Muschinske[3], Bethany M. Niedzielski[4], Kyle Serniak[4], Youngkyu Sung[1,3], Roni Winik[1], Jonilyn L. Yoder[4], Mollie Schwartz[4], Kenji Watanabe[5], Takashi Taniguchi[6], Terry P. Orlando[3], Simon Gustavsson[1], Pablo Jarillo-Herrero[2*], William D. Oliver[1,2,3,4*]

**Affiliations:**

[1]Research Laboratory of Electronics, Massachusetts Institute of Technology, Cambridge, MA 02139, USA.

[2]Department of Physics, Massachusetts Institute of Technology, Cambridge, MA 02139, USA.

[3]Department of Electrical Engineering and Computer Science, Massachusetts Institute of Technology, Cambridge, MA 02139, USA.

[4]MIT Lincoln Laboratory, 244 Wood Street, Lexington, MA 02421, USA.

[5]Research Center for Functional Materials, National Institute for Materials Science, 1-1 Namiki, Tsukuba 305-0044, Japan.

[6]International Center for Materials Nanoarchitectonics, National Institute for Materials Science, 1-1 Namiki, Tsukuba 305-0044, Japan

†These authors contributed equally to this work.

*Correspondence to: joelwang@mit.edu; pjarillo@mit.edu; william.oliver@mit.edu

## Abstract:

**Dielectrics with low loss at microwave frequencies are imperative for high-coherence solid-state quantum computing platforms. We study the dielectric loss of hexagonal boron nitride (hBN) thin films in the microwave regime by measuring the quality factor of parallel-plate capacitors (PPCs) made of $NbSe_2$-hBN-$NbSe_2$ heterostructures integrated into superconducting circuits. The extracted microwave loss tangent of hBN is bounded to be at most in the mid-$10^{-6}$ range in the low temperature, single-photon regime. We integrate hBN PPCs with aluminum Josephson junctions to realize transmon qubits with**

**coherence times reaching 25 μs, consistent with the hBN loss tangent inferred from resonator measurements. The hBN PPC reduces the qubit feature size by approximately two-orders of magnitude compared to conventional all-aluminum coplanar transmons. Our results establish hBN as a promising dielectric for building high-coherence quantum circuits with substantially reduced footprint and, with a high energy participation that helps to reduce unwanted qubit cross-talk.**

**Main Text**

A generalized superconducting qubit comprises Josephson junctions shunted by inductive and capacitive elements that together determine its energy spectrum (*1*). While the materials comprising superconducting qubits would ideally be dissipationless, a dominant contributor to qubit decoherence is the interaction of the electromagnetic fields of the qubit with lossy bulk and interfacial dielectrics (*2*).

In a typical superconducting circuit, dielectric loss may occur in the tunneling barrier of Josephson junctions as well as the native oxide layers covering the many metallic and substrate interfaces of the device (*3*, *4*). These dielectrics are typically amorphous oxides with structural defects that can be modeled as spurious two-level systems (TLSs). Although the microscopic nature of these TLSs remains to be fully understood, it has been established that the interaction between TLS ensembles and the electromagnetic fields in superconducting quantum circuits limit the coherence of qubits and the quality factor of superconducting resonators. It is also suspected that TLSs may be present at interfaces holding chemical residue left from the device fabrication processes (*4*, *5*).

State-of-the-art superconducting qubit technology has managed to mitigate the impact of dielectric loss to a degree through materials science, fabrication engineering, and primarily improved microwave design (*1*). In particular, current qubit designs employ large coplanar (lateral) capacitor pads, which serve to dilute the energy participation of amorphous interfacial dielectrics at the expense of a large form factor for individual qubits and complexities in microwave design. More importantly, the parasitic capacitive coupling between lateral capacitors, which gives rise to stray qubit-qubit coupling, tends to inadvertently increase coherent (spectator) errors (*6*, *7*), inhibiting high-fidelity quantum operations at scale. This design choice reflects the present lack of a suitable low-loss capacitor dielectric needed to make small form-factor, parallel-plate capacitors.

Hexagonal boron nitride (hBN) is a layered van der Waals (vdW) insulator that features chemical inertness, an atomically-flat interface, crystalline structure, and a lack of dangling bonds (*8*, *9*). It provides a pristine environment for constructing devices for quantum transport, nano-photonics, and graphene-based superconducting qubits (*9–11*). hBN is also critical for stacking different vdW materials in any desired order while preserving the functional and structural integrity of each constituent vdW thin film. The resultant devices, called vdW heterostructures (*9*), exhibit structural likeness to epitaxially grown heterostructures and are readily compatible with circuit quantum electrodynamics (cQED) architectures (*11–13*). Although vdW materials and their heterostructures have long been anticipated to be ideal material platforms for building high-coherence quantum devices, there remains a critical – yet missing – piece of information needed for this application: the dielectric loss tangent of hBN in the presence of microwave electromagnetic fields, especially in the GHz regime where superconducting qubits operate. In this work, we directly measure the dielectric loss of hBN in the GHz regime using a superconducting resonator. We then corroborate that result by making a superconducting qubit with a parallel-plate capacitor using hBN dielectric and demonstrate coherence times comparable with qubits fabricated using conventional lateral capacitors, despite being 250 times smaller in size.

We first study the microwave loss of hBN by incorporating hBN into microwave resonators. The intrinsic loss of a dielectric can be characterized by its loss tangent, $\tan\delta = \text{Im}[\epsilon]/\text{Re}[\epsilon] = 1/Q$, where $\epsilon$ is the permittivity of the material, and $Q$ is the material quality factor. Figure 1A shows the schematic of our experimental design comprising lumped-element microwave resonators with a meander inductor and a capacitor that are inductively coupled to a common feedline. The quality factor of hBN can be inferred from the internal quality factor $Q_i$ of the resonators containing hBN via the transmission coefficient $S_{21}$ of the common feedline.

Since the electric field of the resonator driven on resonance may exist in various materials and not solely in the hBN, one must consider the "participation" of each material and its loss. The participation ratio $p_i$ of each material, defined as $p_i = W_{e,i}/W_{e,tot}$ where $W_e$ is the electric field energy, is used to decompose the measured internal quality factor and its associated total loss tangent $\delta_{tot}$ into a weighted-sum of constituent loss tangents:

$$1/Q_i = \tan\delta_{tot} = \sum p_i \tan\delta_i,$$

where $\tan\delta_i$ is the loss tangent for each participating material $i$ weighted by its participation ratio $p_i$.

To study the dielectric loss of hBN, we incorporate hBN crystals into two types of LC resonators: one with hBN laid on top of an inter-digital capacitor (IDC) (Fig. 1B) and one with a parallel plate capacitor (PPC) (Fig. 1C) using hBN as the capacitor dielectric. The participation ratio of hBN in each scheme is calculated using COMSOL Multiphysics (see supplementary material and Fig.S1), which shows that the electric fields are concentrated at the IDC and within the dielectric portion of the PPC. For the IDC resonator, an 80 μm by 80 μm capacitor and a flake thickness of roughly 500 nm yields an hBN participation ratio around 13%. In contrast, with an hBN dielectric thickness of 10 nm and an out-of-plane dielectric constant $\epsilon_\perp$ = 3.76 (*14*), the PPC for the resonator yields a capacitance around 83 fF and an hBN participation ratio around 66%.

Figure 1B shows a mechanically exfoliated hBN flake that is picked-up and transferred onto the IDC of an aluminum (Al) microwave resonator (hBN-Al-IDC in Fig. 1A) using the standard dry-polymer technique (*11*, *15*). The IDC design enables us to probe hBN in a relatively intrinsic or "untouched" state, since there is no further post-transfer fabrication required. On the same chip, another LC resonator, called Al-IDC, with the same nominal resonance frequency ($f_r$=4.63 GHz) as the unloaded (no hBN) hBN-Al-IDC resonator is implemented as a control resonator. The device is cooled and measured in a dilution refrigerator with a base temperature of approximately 10 mK (see supplementary material for our measurement set up).

We investigate the microwave loss of hBN by measuring the transmission ($S_{21}$, Fig. 1A) of a microwave feedline coupled to the test resonators in a hanger geometry using a vector network analyzer (VNA). Figure 2A plots the internal quality factor $Q_i$ of the hBN-Al-IDC (green data points) and Al-IDC (purple data points) resonator as a function of microwave power. The $Q_i$ is extracted by fitting the line shape of the $S_{21}$ signal around the resonance frequency using a numerical fit that accounts for both the internal and external $Q$ factors ($Q_i$ and $Q_e$ in Fig.1A) (*16*).

In the low temperature regime ($kT \ll \hbar\omega$, where $k$ is the Boltzmann constant, $T$ is the temperature, and $\omega$ is the driving-field frequency), the $Q_i$ of the LC resonator generally increases with driving-field power, which serves to drive TLSs into saturation. Once saturated, the TLSs no longer contribute to the dielectric loss, but simply exchange energy with the resonator electromagnetic field (*2*, *4*). In contrast, the $Q_i$ is generally lowest at the lowest driving powers, where most TLSs are in their ground state and readily able to absorb energy from the resonator electromagnetic field. At the single photon power level (dashed line), the hBN-coupled IDC shows a quality factor of approximately $6\times10^5$, whereas the $Q_i$ of the control IDC is similar or even slightly lower (while fluctuating). Across all of the IDC devices measured, we observe that the presence of hBN does not limit the $Q_i$ of the resonators.

To increase the participation ratio, we perform the same measurements on resonators comprising parallel-plate capacitors formed from hBN dielectric and $NbSe_2$, a vdW superconductor (*17*, *18*) used for the plates. The devices are fabricated as $NbSe_2$-hBN-$NbSe_2$ heterostructures (Fig. 1C). The heterostructure is assembled using dry-polymer techniques in an Ar-filled glovebox to prevent oxidation at $NbSe_2$-hBN interfaces, which enables us to characterize the dielectric loss primarily given by hBN.

The quality factors $Q_i$ versus drive power extracted from three representative PPCs are plotted in Fig. 2B-2D ($f_r$=4.18 GHz, 6.82 GHz, and 10.0 GHz respectively). The $p_{hBN}$ of these devices range from 59% to 67% and is diluted primarily by the parasitic capacitance in the meander inductor, which retains some of the electric field.

Around the single-photon power level, the $Q_i$'s of the PPC resonators (Fig. 2B-2D) are approximately $2.3 \times 10^5$, $3.4 \times 10^5$, and $1.9 \times 10^5$, respectively. We average the first 5 data points below the single photon limit to reduce the influence of outliers. Above the single-photon limit, the $Q_i$ generally increases with drive power until it declines at the higher power regime (at around -110 dBm). This behavior is reminiscent of observations in Al resonators attributed to a non-equilibrium distribution of quasiparticles induced at high microwave drive powers that results in an additional power-dependent loss channel (*19*). Given the higher participation of the hBN in the PPC configuration, a more accurate measure of the hBN material quality factor can be extracted. We estimate the hBN loss tangent $\tan \delta_{hBN}$ by decomposing the total loss as $\tan \delta_{tot} = 1/Q_i = p_{hBN} \tan \delta_{hBN} + p_r \tan \delta_r$, where $p_r$ and $\tan \delta_r$ denotes the participation ratio and loss tangent of the rest of the circuit respectively. To calculate the worst-case upper bound of $\tan \delta_{hBN}$, we can attribute all of the loss to the hBN, *i.e.*, by assuming $\tan \delta_r = 0$. The resulting upper bound for $\tan \delta_{hBN}$ is obtained by dividing $\tan \delta_{tot}$ by $p_{hBN}$, which yields $6.7 \times 10^{-6}$, $5.0 \times 10^{-6}$, and $7.8 \times 10^{-6}$, respectively. This directly translates to a worst-case lower bound for the hBN quality factor: $Q_{hBN} \geq 1.5 \times 10^5, 2.0 \times 10^5, 1.3 \times 10^5$ respectively. Across all eleven measured PPC devices, eight exhibit quality factors $Q_i = 2\pi f_{01} T_1$ between $1 \times 10^5$ and $3 \times 10^5$, corresponding to an energy relaxation time $T_1$ between 3.9 μs and 12 μs, assuming a resonance frequency $f_{01}$ = 4 GHz for the resonators.

hBN as a low-loss dielectric offers a promising opportunity to construct high-coherence superconducting qubits with small layout geometries. A PPC with low-loss hBN dielectric features high specific capacitance (capacitance per unit area) and maintains the capacitor electric field within a small volume consisting of high-quality hBN dielectric. Figure 3A shows such a circuit with three PPC-shunted transmons and one conventional transmon qubit with an aluminum lateral capacitor, each coupled to an individual readout resonator and sharing a common microwave feedline. The PPC-shunted transmons are constructed on a silicon chip with pre-fabricated aluminum circuit elements, including the transmission line, readout resonators, Josephson tunnel junctions, and the aluminum transmon (Fig. 3A-3C). The NbSe$_2$-hBN-NbSe$_2$ capacitors are then integrated into the circuit using the same approach as the resonator devices. The target capacitance of each PPC is 90 fF, with hBN thickness ranging from 10 nm to 30 nm for the corresponding plate areas spanning 27 μm$^2$ to 80 μm$^2$. The hBN PPC is about 250 times smaller than the lateral capacitor of the aluminum transmon that dominates the footprint of typical Xmon-type geometry (*20*), while achieving energy participation ranging from 76.8% ~ 91.0% (Table 1).

We perform time-domain measurements of fixed-frequency and flux-tunable PPC-shunted transmons in the same measurement setup as the resonator measurements. The fixed-frequency design allows us to probe the microwave loss of hBN-based capacitors independent of the magnetic flux-noise in the system; whereas the tunable-frequency design enables us to assess dephasing noise due to the introduced susceptibility to magnetic flux noise. Although 1/f magnetic flux noise is generally associated with dephasing, noise at the qubit frequency may also contribute to energy decay (loss) (*21*).

Figure 4A plots the excited-state probability of a fixed-frequency PPC-shunted transmon (device ID: NT-PPCQ4) initially prepared in its excited state as a function of pulse delay $\tau_{\text{delay}}$, from which we extract the energy relaxation time $T_1$ = 24.4 μs. In Fig. 4B, we show Ramsey interference fringes, from which we obtain the decoherence time $T_2^*$ = 25.1 μs (Fig. 4B). In addition, this device has a mean $T_1$ of 17.9 μs and a mean $T_2^*$ of 17.2 μs when averaged over 64 separate measurements within a 12-hour window (Fig. 4A, 4B histograms). The coherence times of this best-performing fixed-frequency PPC-shunted qubit are approximately within a factor 2-3 of those from the fixed-frequency aluminum transmon on the same chip, with a mean (best) $T_1$ = 49.0 (51.6) μs and $T_2^*$ = 38.2 (73.6) μs. We also characterize $T_1$ as a function of qubit transition frequency $f_{01}$ of a flux-tunable PPC-shunted transmon (Fig. 4C, device ID: T-PPCQ1). As shown in figure 4D, the $T_1$ of such flux-tunable qubits tends to grow, albeit with temporal fluctuations, as the $f_{01}$ decreases, a trend consistent with the qubit coupling capacitively to the dielectric and Fermi's golden rule (*22*).

The coherence times and relevant parameters for both fixed-frequency and flux-tunable PPC-shunted qubits are summarized on Table 1. We emphasize that the coherence times of the best performing PPC-shunted qubits, for either fixed-frequency of flux-tunable designs, are comparable to the standard aluminium qubit on the same chip while being more than 250 times smaller in size. The mean quality factors of the PPC-shunted qubits, estimated by presuming the slightly anharmonic qubits are harmonic oscillators, ranges from $4.2 \times 10^4$ to $4.5 \times 10^5$ and is consistent with the PPC quality factor characterized in the LC resonator (Fig. 3).

The device-to-device variation in coherence times in superconducting qubit devices is typically attributed to TLSs residing in circuit elements such as Josephson junctions and shunting capacitors. For hBN PPCs, both the materials and the associated fabrication processes could introduce TLSs into the heterostructures. For example, the dry-polymer-based stacking approach may form randomly distributed pockets filled with air or hydrocarbons between different layers (*23*). These trapped materials, along with deformations of the layered materials, give rise to structural inhomogeneity that would enhance the dielectric loss in regions with high energy participation . In addition, the entire fabrication process that incorporates the vdW heterostructures in to the superconducting circuit involves a series of thermal cycling and nanofabrication steps, which may undermine the quality factor of pre-fabricated elements such as Josephson junctions and resonators. It is therefore conceivable that our range of extracted hBN quality factor values – already a worst-case lower bound due to the assumption that all observed loss came from the capacitor – should be furthermore treated as a lower bound for the intrinsic hBN crystals and associated vdW-heterostructures. Moreover, we observe no evident correlations between $Q_i$ values and dimensional parameters (the hBN thickness and the PPC area) from measuring 16 PPC devices (11 in LC resonators and 5 in the PPC-shunted transmon qubits) in this experiment (see Table 1 and Fig. S3 in the supplementary information). We attribute this feature and the lower loss as compared to amorphous BN (*24*) to the highly-crystalline bulk materials and atomically flat interfaces in the $NbSe_2$-hBN-$NbSe_2$ heterostructures. One expects to improve the coherence of PPC-shunted qubits or other vdW-based qubits by engineering the fabrication steps to reduce their impact on the materials.

In summary, we have characterized the microwave loss of hBN using resonators and qubits comprising hBN capacitors. The $NbSe_2$-hBN-$NbSe_2$ heterostructure, operated as a parallel-plate capacitor in a superconducting LC circuit, exhibits an intrinsic quality factor up to $3.4 \times 10^5$ at the single photon limit, indicating comparable or even better performance than epitaxially grown lumped-element devices reported to date (*25–27*). The transmon qubits made with hBN PPC's exhibit coherence times up to 25 μs, comparable to an aluminum Xmon qubit fabricated on the same chip. Our results demonstrate that hBN is a relatively high-quality, low-loss dielectric that can be employed to build high-coherence quantum devices in the circuit quantum electrodynamics (cQED) architecture. In addition, qubits shunted with hBN PPCs feature a single-qubit footprint smaller by at least two orders of magnitude than conventional lateral capacitor designs, while containing up to 91% of electric field energy between the parallel plates. We emphasize that this demonstration suggests that high-quality, lumped-element devices may be built with vdW heterostructures to mitigate unwanted qubit-qubit cross-talk, which is key to realizing high-fidelity, multi-qubit operations. Of particular interest is a vdW heterostructure comprising a Josephson tunnel junction and a shunting capacitor using hBN dielectric. Such vdW-based merged-element devices possess the potential to yield high-coherence qubits with an even further reduced form factor (*28*, *29*). Finally, with advancements in wafer-scale growth of single-crystal hBN thin films (*30*, *31*) and vdW heterostructures, one can envision using standard wafer-scale fabrication approaches to manufacture devices incorporating hBN as a substrate, a passivation layer, as lumped-elements like parallel-plate capacitors or Josephson junctions, and even all-vdW materials platforms for extensible superconducting quantum computing schemes.

We became aware of a complimentary work (*32*) during the preparation of this manuscript.

**Acknowledgements**

We acknowledge helpful discussions with Greg Calusine, Thomas Hazard, Dahlia Klein, David MacNeill, Kevin O'Brien, Agustin Di Paolo, and Antti Vepsäläinen. The authors thank Rabindra Das at MIT Lincoln Laboratory for technical assistance.

This research was funded in part by the US Army Research Office grant no. W911NF-18-S-0116, by the National Science Foundation QII-TAQS grant no. OMA-1936263, and by the Assistant Secretary of Defense for Research & Engineering via MIT Lincoln Laboratory under Air Force contract no. FA8721-05-C-0002. K.W. and T.T. acknowledge support from the Elemental Strategy Initiative conducted by the MEXT, Japan (Grant Number JPMXP0112101001) and JSPS

KAKENHI (Grant Numbers 19H05790 and JP20H00354). The views and conclusions contained herein are those of the authors and should not be interpreted as necessarily representing the official policies or endorsements of the US Government.

**Author Contributions**

**Methods**

Details about device fabrication, measurement setups and simulations are available in the supplementary information of this paper. The data that support the findings of this study are available from the corresponding author upon reasonable request and with the cognizance of our US Government sponsors who funded the work.

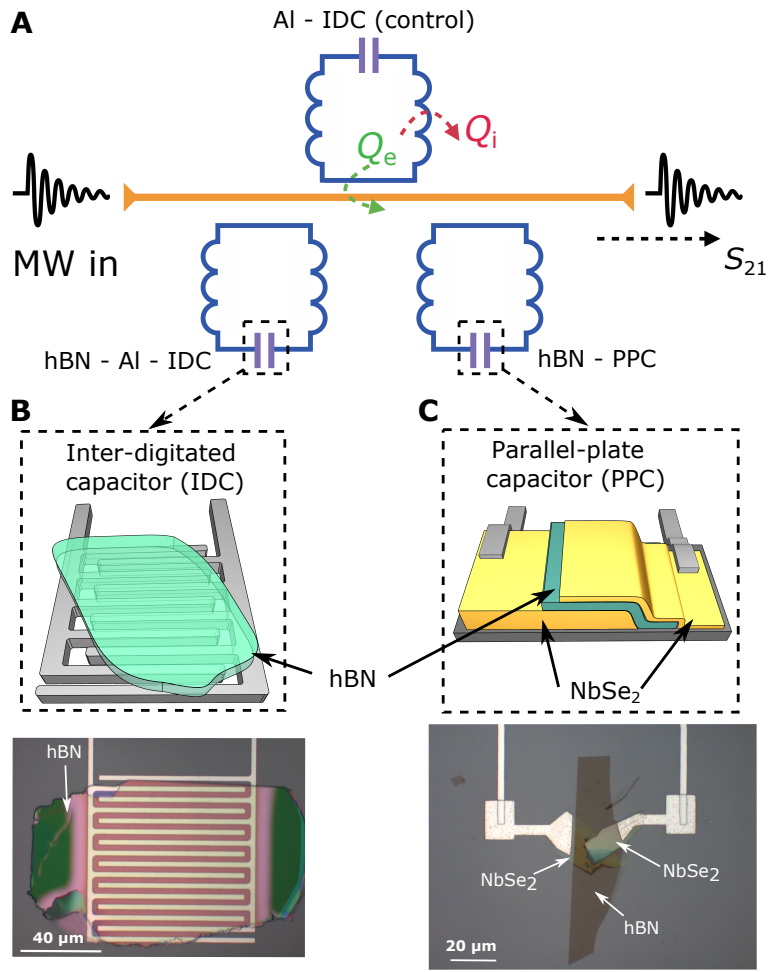

**Fig. 1. Superconducting resonators for characterizing the microwave dielectric loss of hBN.** (**A**) Schematic of the device design. A common transmission line couples inductively to three types of LC resonators, including a control resonator with aluminium inter-digitated capacitor (IDC). The external (coupling) quality factor $Q_e$ and intrinsic quality factor $Q_i$ that depends on the microwave loss properties of hBN are extracted from the $S_{21}$ signal measured by a vector network analyzer (VNA). (**B**) hBN-coupled IDC to study the microwave loss of hBN. The hBN flake is picked-up and transferred onto the IDC, without any subsequent fabrication steps. (**C**) LC resonator consisting of a NbSe$_2$-hBN-NbSe$_2$ vdW heterostructure serving as a parallel-plate capacitor.

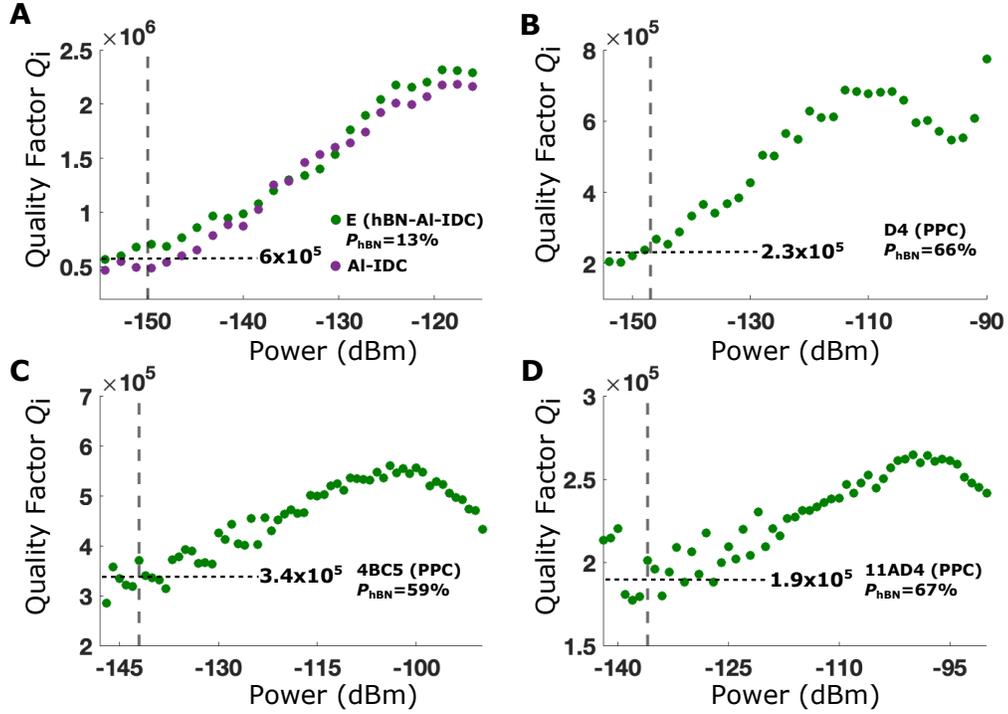

**Fig. 2. Internal quality factor $Q_i$ of hBN-coupled LC resonators.** (**A**) $Q_i$'s of LC resonators with hBN-covered IDC (Fig. 1(B), green data points) and the control IDC (purple data points) plotted as a function of microwave power. The quality factors are obtained by fitting the line shape of the transmitted signal ($S_{21}$) around the resonance frequency (*16*). At the single-photon power level (vertical dashed line), the two resonators exhibit similar $Q_i$. (**B**), (**C**) and (**D**) $Q_i$'s of three different resonators with NbSe$_2$-hBN-NbSe$_2$ PPCs, all showing $Q_i$ greater than $2\times 10^5$ at the single-photon limit (vertical dashed lines). The decrease of $Q_i$ at higher powers (> -100 dBm) may be attributed to a non-equilibrium distribution of quasiparticles induced by high microwave power and effective electron temperature.

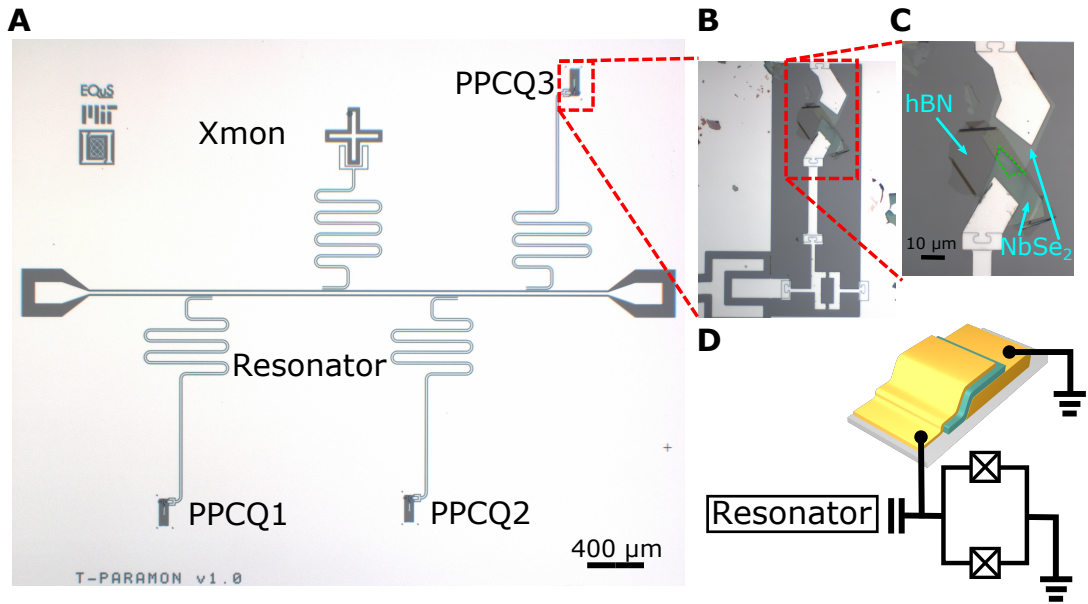

**Fig. 3. PPC-shunted transmon qubit device.** (**A**) Optical image of a PPC-shunted transmon chip. Four qubits, including 3 PPC-shunted transmon qubits (PPCQ) and one standard Xmon qubit, share a common line for readout and qubit control. All aluminum parts, except for the bridging electrodes connecting the PPC to the rest of circuit, are pre-fabricated before transferring the vdW heterostructures. (**B**) Zoom-in image of the PPC-shunted transmon qubit, showing the SQUID made of Al/AlO$_X$/Al tunnel junctions. (**C**) PPC used in the PPC-shunted transmon qubit. The capacitance is defined by the overlap region ($\approx$ 54 μm$^2$) marked by the dashed line. (**D**) Schematic of the flux-tunable PPC-shunted transmon qubit.

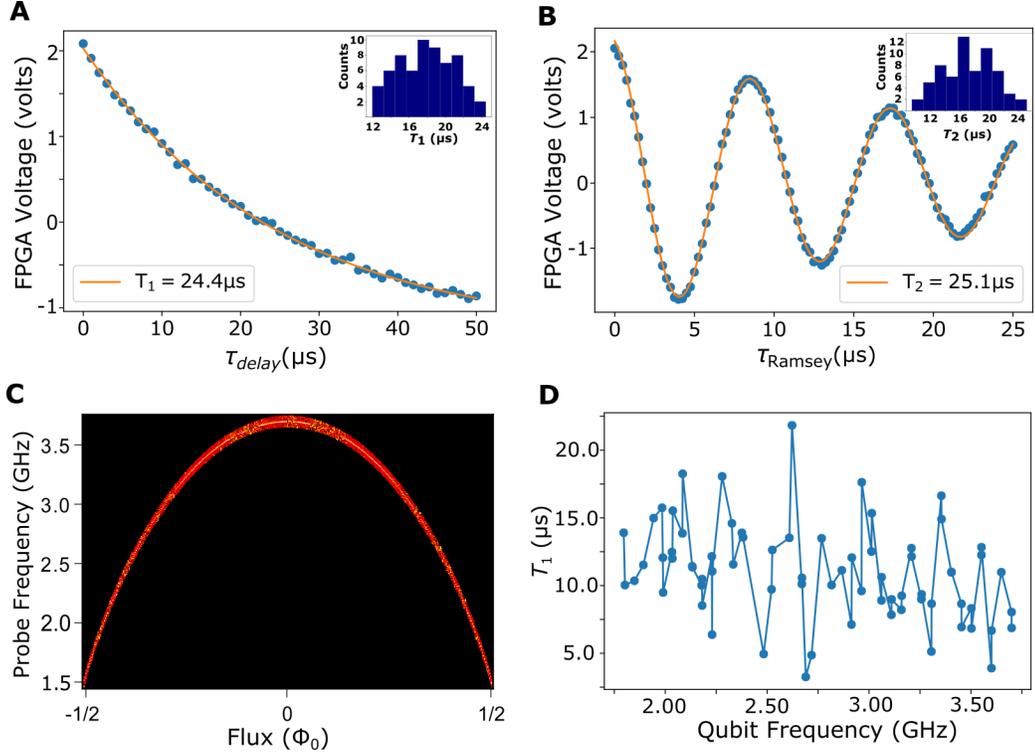

**Fig. 4. Characterization of fixed-frequency and flux-tunable PPC-shunted transmon qubits.** (**A**) Energy relaxation time $T_1$ measurement of a fixed-frequency PPC-shunted transmon device (NT-PPCQ3). The fitting of the qubit population to an exponential decay function yields $T_1 \sim 24.4$ μs. Inset: histogram from 64 averaged measurements taken over 12 hours with $T_1$ ranging between 12~24 μs. (**B**) Ramsey measurement of a fixed-frequency PPC-shunted transmon device (NT-PPCQ3). The dephasing time $T_2^*$ of 25.1 μs is obtained by fitting to the function $\exp(-\tau_{Ramsey}/T_2^*) \times \cos(2\pi\Delta f \times \tau_{Ramsey})$. Inset: histogram from 64 averaged measurements taken over 12 hours with $T_2^*$ ranging between 10~24 μs. (**C**) Energy spectrum of a flux-tunable PPC-shunted transmon device (T-PPCQ1). (**D**) $T_1$ of a flux-tunable PPC-shunted transmon device (T-PPCQ1) measured as a function of qubit frequency. The $T_1$ shows an ascending trend as the qubit frequency decreases.

| Device ID | $f_{01}$ (GHz) | $\alpha/2\pi$ (MHz) | $T_1$ (μs) Best | $T_1$ (μs) Mean | $T_1$ (μs) Std. | Mean Q ($10^3$) | $T_2^*$ (μs) Best | $T_2^*$ (μs) Mean | $T_2^*$ (μs) Std. | $p_{hBN}$ (%) | $t_{hBN}$ (nm) | $A_{PPC}$ (μm²) |
|---|---|---|---|---|---|---|---|---|---|---|---|---|
| NT-Xmon | 3.69 | 222.2 | 51.6 | 49.0 | 1.70 | - | 73.6 | 38.2 | 13.8 | - | - | - |
| NT-PPCQ1 | 3.61 | 206.6 | 22.4 | 14.9 | 2.84 | 338.2 | 21.2 | 9.1 | 5.12 | 76.8 | 30 | 97.0 |
| NT-PPCQ2 | 3.50 | 154 | 4.6 | 2.0 | 0.84 | 43.4 | 4.3 | 2.7 | 0.67 | 91.0 | 10 | 174.9 |
| NT-PPCQ3‡ | 3.99 | 191.2 | 24.4 | 17.9 | 3.03 | 448.6 | 25.1 | 17.2 | 3.55 | 82.7 | 21 | 75.3 |
| T-Xmon | 3.80 | 212.6 | 52.9 | 45.5 | 3.44 | - | 5.36 | 4.93 | 0.16 | - | - | - |
| T-PPCQ1‡ | 3.82 | 230 | 16.1 | 12.1 | 1.70 | 290.6 | 17.1 | 8.34 | 2.61 | 81.0 | 24 | 68.0 |
| T-PPCQ2 | 3.47 | 174 | 3.65 | 1.94 | 0.46 | 42.3 | 5.50 | 2.67 | 0.76 | 82.8 | 21 | 78.5 |

**Table 1: Characteristics of transmon qubits shunted by parallel-plate capacitors (PPC) and conventional Xmon qubits.** Upper panel: fixed-frequency transmons. Lower panel: flux-tunable transmons. ‡ denotes fixed-frequency and tunable transmons discussed in the main text. The thickness of hBN ($t_{hBN}$) is obtained by atomic force microscopy (AFM), and the PPC area ($A_{ppc}$) is defined by the overlapping region of each NbSe$_2$-hBN-NbSe$_2$ stack.

Supplementary Information for

# Hexagonal Boron Nitride (hBN) as a Low-loss Dielectric for Superconducting Quantum Circuits and Qubits

Joel I-J. Wang[1†*], Megan A. Yamoah[2,3†], Qing Li[3], Amir H. Karamlou[1,3], Thao Dinh[2], Bharath Kannan[1,3], Jochen Braumueller[1], David Kim[4], Alexander J. Melville[4], Sarah E. Muschinske[3], Bethany M. Niedzielski[4], Kyle Serniak[4], Youngkyu Song[1,3], Roni Winik[1], Jonilyn L. Yoder[4], Mollie Schwartz[4], Kenji Watanabe[5], Takashi Taniguchi[6], Terry P. Orlando[3], Simon Gustavsson[1], Pablo Jarillo-Herrero[2*], William D. Oliver[1,2,3,4*]

[1]Research Laboratory of Electronics, Massachusetts Institute of Technology, Cambridge, MA 02139, USA.

[2]Department of Physics, Massachusetts Institute of Technology, Cambridge, MA 02139, USA.

[3]Department of Electrical Engineering and Computer Science, Massachusetts Institute of Technology, Cambridge, MA 02139, USA.

[4]MIT Lincoln Laboratory, 244 Wood Street, Lexington, MA 02421, USA.

[5]Research Center for Functional Materials, National Institute for Materials Science, 1-1 Namiki, Tsukuba 305-0044, Japan.

[6]International Center for Materials Nanoarchitectonics, National Institute for Materials Science, 1-1 Namiki, Tsukuba 305-0044, Japan.

†These authors contributed equally to this work.

*Correspondence to: joelwang@mit.edu; pjarillo@mit.edu; william.oliver@mit.edu

## 1 Device Fabrication

All devices are fabricated from a 250 nm-thick aluminum film deposited on highly-resistive Si chips. The chips for hBN-Q devices, including the IDC and PPC resonators (Fig. 1A and 1B in the main text), are patterned with direct-write photolithography technique using maskless aligner (Heidelberg MLA150), whereas the PPC-shunted transmon qubit chips are fabricated using standard photolithography techniques. Josephson junctions used in conjunction with the vdW materials capacitor are fabricated by double-angled shadow evaporation of aluminium. Details of the aluminium fabrication process can be found in Ref (*1*).

The incorporation of vdW heterostructures onto the pre-patterned chips employs standard mechanical exfoliation and dry polymer-based techniques (*2*, *3*). The NbSe$_2$ flakes used for PPC and PPC-shunted transmon devices are exfoliated and assembled with hBN to form the NbSe$_2$-hBN-NbSe$_2$ heterostructures in an Argon-filled dry box.

After transferring the heterostructures, the bridging between NbSe$_2$ to the pre-fabricated aluminum elements such as meander inductor, ground plane, and the electrode of the Josephson junction, is done by in-situ ion-mill and aluminum deposition.

## 2 Microwave simulation

To determine the participation ratios of electric field energy in each material of our device, we use COMSOL Multiphysics finite element solver. We model the device conductor as a two-dimensional perfect electrical conductor with a 279 μm Silicon (Si) substrate and 800 μm of vacuum above and below the model and calculate the eigenmodes of the system. The first simulation of the hBN-IDC-Control resonator, without an hBN flake, allows us to determine the participation of the Si substrate in for comparison. Each hBN-IDC device we fabricate has varying thicknesses and sizes of hBN flakes. As such, each is simulated to determine the participation of the hBN and the Si substrate.

Finally, due to the difficulties of meshing irregular shapes and across dimensions of four orders of magnitude, we make approximations to the hBN flake shape as shown in Fig. S1.

## 3 Measurement Setup

The experiment is performed in a Bluefors XLD-1000 dilution refrigerator with a base temperature of ≈10 mK. All attenuators in the cryogenic samples are made by XMA and installed to remove excess thermal photons from higher-temperature stages. We pump the Josephson travelling wave parametric amplifier (TWPA) to pre-amplify the readout signal at base temperature. To avoid any back- action of the pump-signal from TWPA, we add a microwave isolator between the samples and the TWPA. On the RF output line, there is a high-electron mobility transistor (HEMT) amplifier (by LNF) thermally connected to the 3 K stage. At the room temperature (300K) stage, we implement another amplifier (MITEQ) to further enhance the output signal. See Fig. S2 for the schematics of the measurement setup.

## 4 Additional PPC samples

In addition to the PPC devices discussed in the main text (Fig.2 B, C, and D), we present the measurement results from 8 additional PPC resonator devices in the same setup as summarized in the table below.

| Device ID | $Q_i$ ($10^3$) | $T_1$ (μs) | $t_{hBN}$ (nm) | $A_{PPC}$ (μm²) | $p_{hBN}$ (%) |
|---|---|---|---|---|---|
| V4-F6-L | 147.8 | 3.9 | 30 | 77.7 | 57.0 |
| V4-F6-R | 105.2 | 4.7 | 21 | 91.0 | 69.0 |
| V3-E4 | 81.3 | 2.7 | 10 | 28.1 | 64.2 |
| V4-1A | 176.5 | 3.3 | 22 | 74.0 | 64.0 |
| V4-B5 | 256.3 | 5.0 | 26 | 140.0 | 74.6 |
| V4-B3 | 323.3 | 6.4 | 31 | 51.4 | 44.4 |
| V3-A2 | 171.8 | 4.8 | 10 | 7.9 | 36.5 |
| V4-5B | 88.1 | 1.4 | 35 | 28.4 | 32.6 |
| V4-17A-2 | 73.7 | 2.5 | 28 | 40 | 46.2 |

**Table S1: Summary of quality factor measurements from additional PPC samples.** $T_1$ is calculated using the relation $Q_i = 2\pi f T_1$ for a resonator. PPC area is defined by the overlapping region of the NbSe$_2$- hBN-NbSe$_2$ stack.

## 5 Correlations between $Q_i$ and PPC dimensional parameters

Figure S3 shows the plots of $Q_i$ versus dimensional parameters including hBN thickness, PPC area and the hBN participation ratio ($p_{hBN}$). Measurement of the $Q_i$ is performed over 5 PPC-shunted transmon qubits (PPC-Qubit), 3 PPCs discussed in the main text (PPC-MT), and the 8 additional PPCs (PPC-SI) shown in section 4.

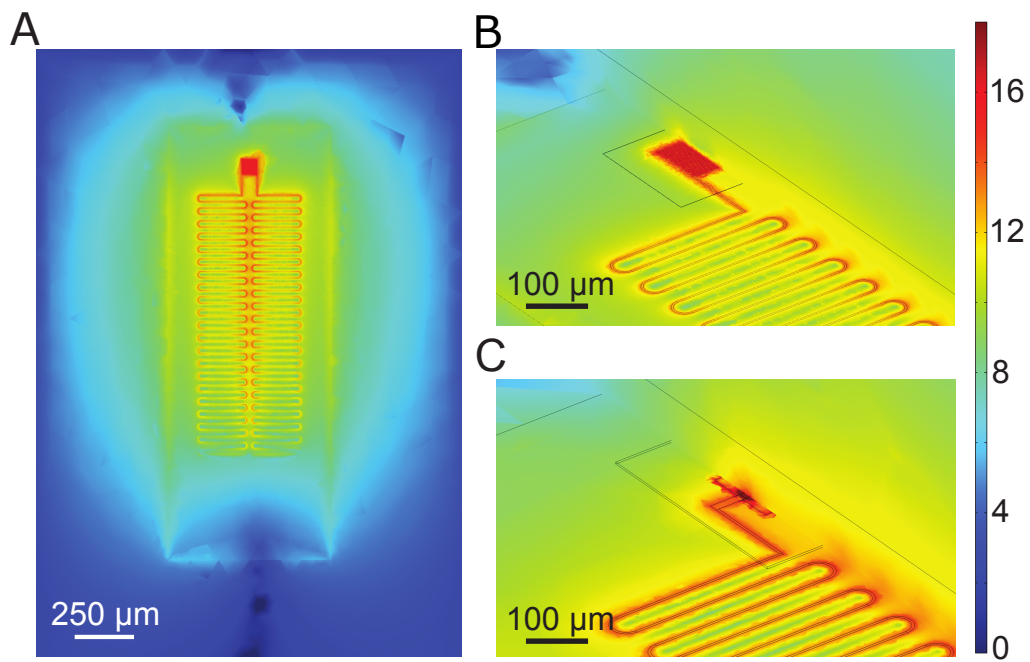

**Fig. S1. Electric field energy distribution of resonator devices on resonance as simulated in COMSOL Multiphysics.** (**A**) Simulation of hBN-IDC-Control device with IDC capacitor not coupled to an hBN flake. (**B**) Close-up of capacitor for hBN-IDC device in (**A**) but with an hBN flake coupled to the IDC. (**C**) Close-up on PPC of the hBN-PPC device. In all, electric field energy plotted in logarithmic scale.

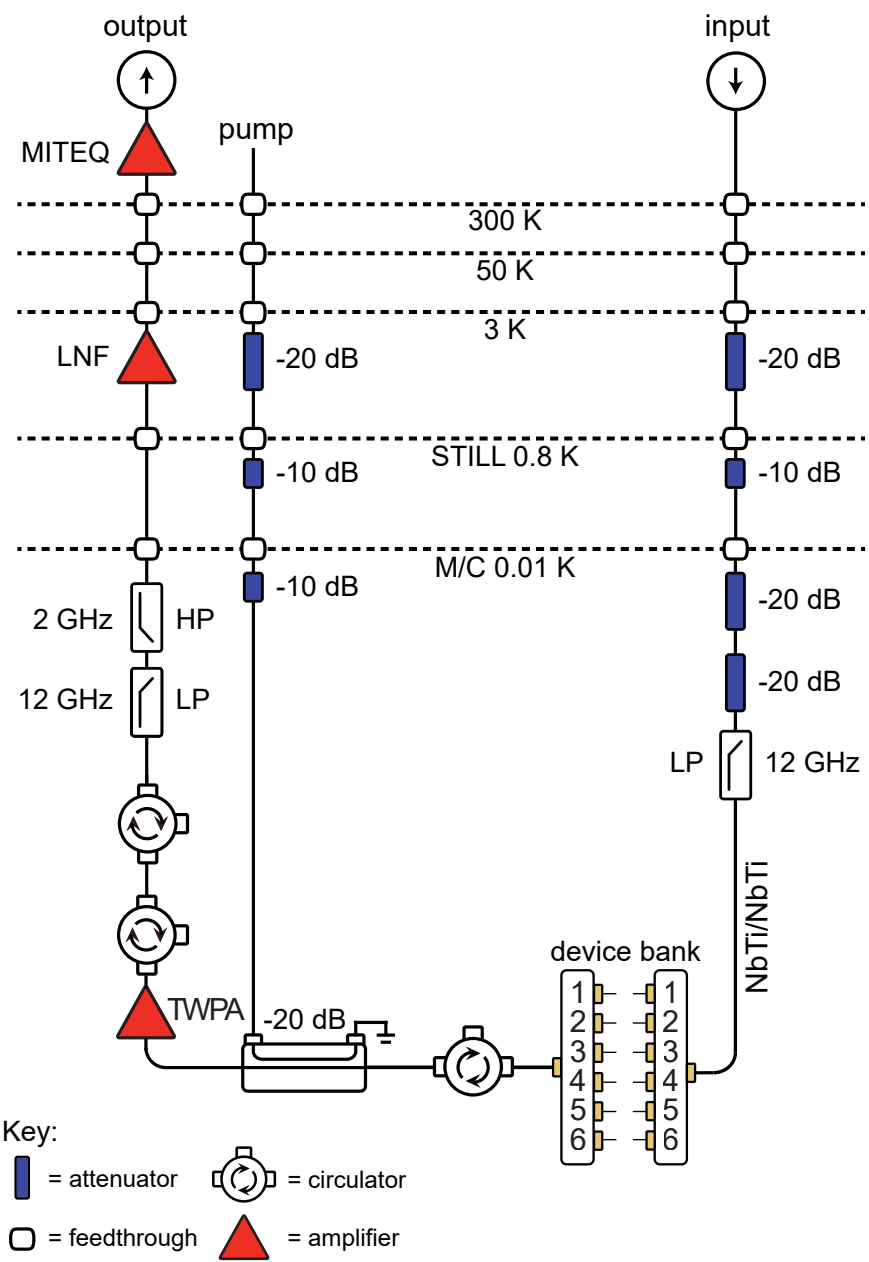

**Fig. S2. Schematics of the measurement setup.**

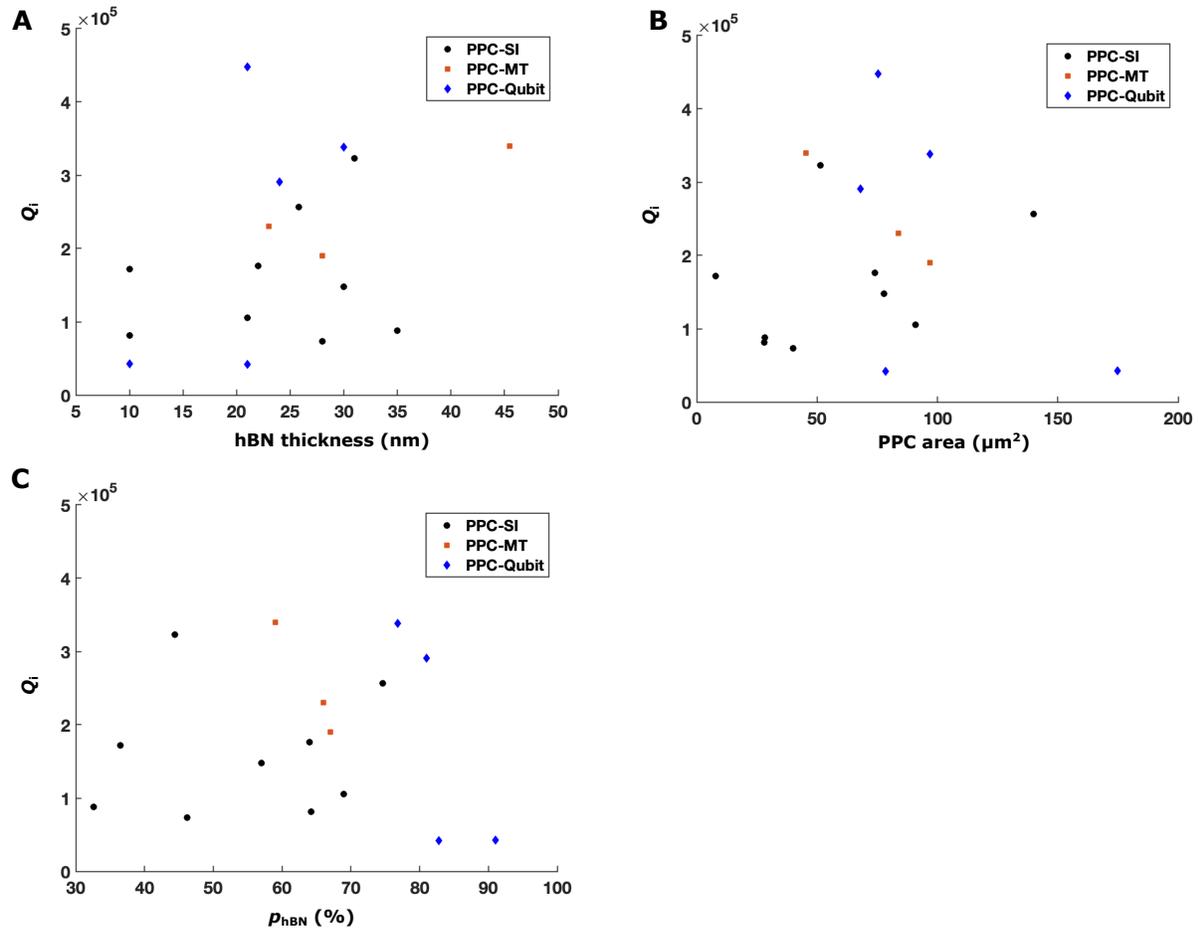

**Fig. S3. Internal quality factors from 16 hBN PPCs of various sizes.** (**A**) Plot of internal quality factors versus hBN thicknesses. (**B**) Plot of internal quality factors versus the overlapping area of the $NbSe_2$- hBN-$NbSe_2$ stack. (**C**) Plot of internal quality factors versus hBN participation ratios. The 16 devices include PPCs used in the PPC-shunted transmon qubits (PPC-Qubit, blue diamond), PPCs discussed in the main text (PPC-MT, orange square), and the additional PPCs shown in the supplementary information (PPC-SI, black circle).

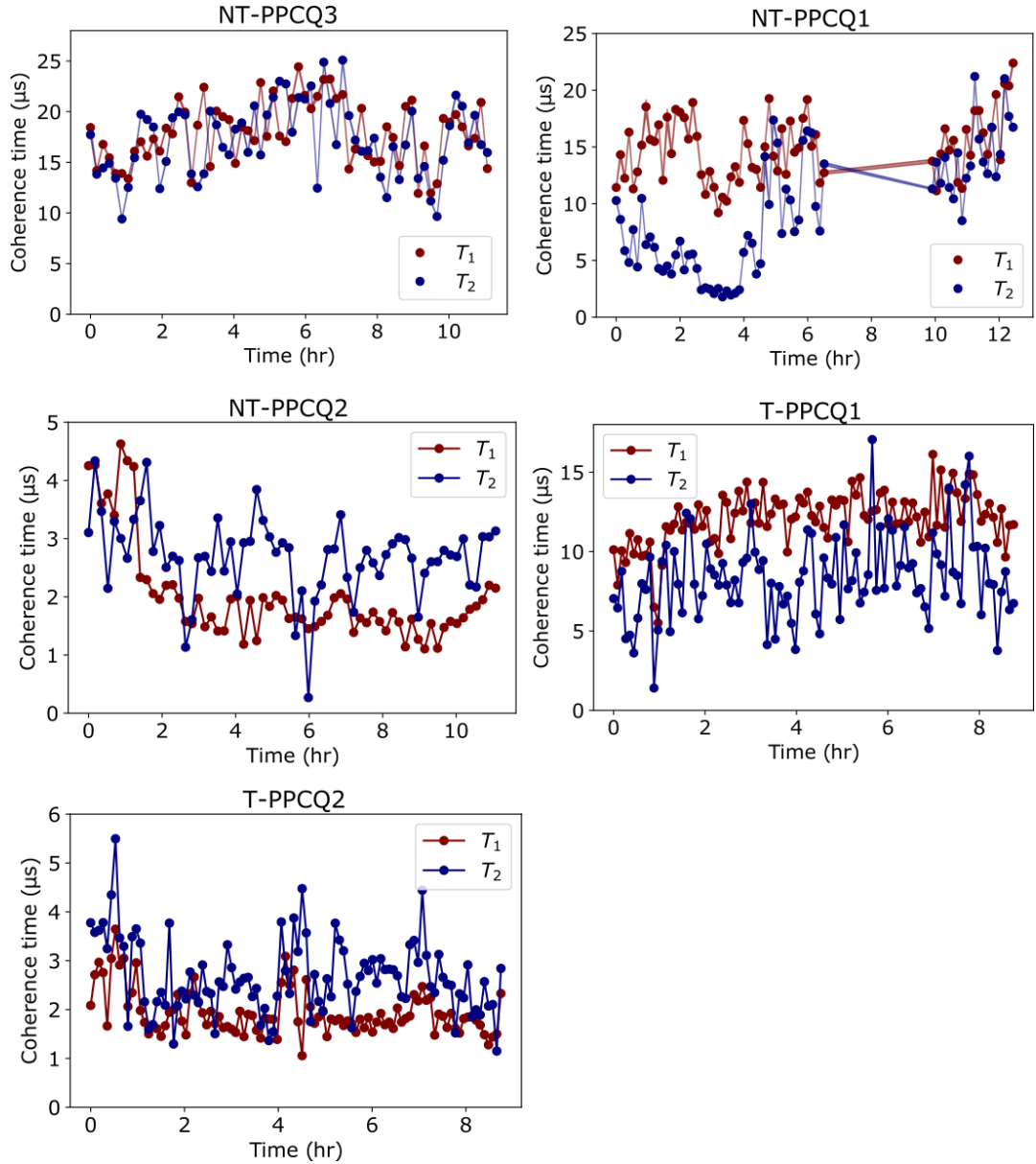

**Fig. S4. Coherence times of PPC-shunted transmon qubits taken over 12 (NT-PPCQ1, NT-PPCQ2, and NT-PPCQ3) and 10 (T-PPCQ1 and T-PPCQ12) hours.**